\documentclass[onecolumn,showpacs,preprintnumbers,nofootinbib,amsmath,amssymb]{revtex4}

\usepackage{amsfonts} 
\usepackage{amssymb}  
\usepackage{graphicx} 
\usepackage{amsmath}  
\usepackage{amsmath}  
\usepackage[latin1]{inputenc}
\usepackage{color}

\usepackage{aas_macros}
\usepackage{natbib}

\begin{document}
\title{Magnetized BEC stars with boundary conditions depending on magnetic field}
\author{G. Quintero Angulo\footnote{gquintero@fisica.uh.cu}}
\affiliation{Facultad de F{\'i}sica, Universidad de la Habana,\\ San L{\'a}zaro y L, Vedado, La Habana 10400, Cuba}
\author{A. P\'erez Mart\'{\i}nez\footnote{aurora@icimaf.cu}}
\affiliation{Instituto de Cibern\'{e}tica, Matem\'{a}tica y F\'{\i}sica (ICIMAF), \\
 Calle E esq a 15 Vedado 10400 La Habana Cuba\\}
\author{H. P\'erez Rojas\footnote{hugo@icimaf.cu}}
\affiliation {Instituto de Cibern\'{e}tica, Matem\'{a}tica y F\'{\i}sica (ICIMAF), \\
 Calle E esq a 15 Vedado 10400 La Habana Cuba\\}

\thanks{}%
\date{\today}%

\begin{abstract}
 In this paper we analyze the impact of the boundary conditions on the solutions of the $\gamma$-structure equations for magnetized Bose--Einstein condensate stars (BECs). We consider two cases: a pure BECs and a BECs embedded in a magnetic field. Since they correspond to different physical situations, they lead to different numerical behavior, having important effects on the stability of the solutions as well as on the observables of the star. The differences between the mass--radii curves obtained in each situation are presented and discussed.

\end{abstract}


\maketitle

\section{Introduction}\label{sec1}

Compact objects are explained by quantum principles. The most common is to suppose that they are composed of degenerated gases of fermions, whose Fermi pressure --a consequence of the Pauli exclusion principle--, counteracts the collapse. But a degenerated boson gas, despite the fact that its pressure is very small due to Bose-Einstein condensation, can also counterbalance gravity due, ultimately, to Heisenberg's uncertainty principle. 

First theoretical works related to stars fully composed of non-interacting bosons appeared in the sixties of the last century \cite{PhysRev.187.1767}. Nevertheless, these models were unattractive from the compact objects point of view, since they give extremely light objects whose properties cannot be connected to any observed star. Typical masses of fermions stars scale as $M_{crit}\sim\frac{M_{PL}^3}{m^2}$\footnote{  $M_{PL}=\sqrt{\frac{hc}{G}}$ is the well-known Plank mass that depends on universal constants.}, while for bosons $M_{crit}\sim\frac{M_{PL}^2}{m}$, being $m$ the mass of the individual particles \cite{PhysRevD.38.2376}. For fermions (bosons) stars make of particles with the nucleon mass these expressions give $M_{crit}\sim0.7 M_{\odot}$ ($M_{crit}\sim10^{-20}M_{\odot}$). Maximum masses of boson stars can be increased by adding interactions between the particles \cite{Takasugi1984}. However, those stars models should wait until the beginning of the new millennium to have an astrophysical revival led by the achievement of the BEC \cite{Anderson1995} and the BEC-BCS crossover in lab \cite{Randeria,Leggett,Parish}, besides the adjustment of the observational cooling data of the neutron star of Cassiopeia A \cite{Shternin:2010qi,Page:2011yz,PageSC}.

The combination of those three facts have suggested that neutrons inside neutron stars (NS) cores might be forming spin one pairs that, under certain conditions, behave as effective vector bosons (see. Refs. \cite{Chavanis2012,Angulo:2018url} for a more detailed explanation). Therefore, at some stage of their evolution, the inner shells of NS could be fully or mostly composed by this matter, an statement that give birth to the Bose--Einsten condensate stars (BECs) as an alternative model to NS core. A BECs is a star fully composed of interacting bosons formed by the spin parallel pairing of two neutrons. These particles counterbalance gravity with the pressure that comes from their interactions, whose strength determine the maximum mass of the compact object, that can be as high as the nowadays desirable two solar masses \cite{Chavanis2012,latifah2014bosons}.

Since neutrons stars are well known for being strongly magnetized objects \cite{Lattimerprognosis}, in a previous work we incorporate magnetic field to the BECs description \cite{Angulo:2018url}. In our model, magnetic field is always  in the $z$ direction, $\textbf{B}=(0,0,B)$, and its intensity $B$ might be constant or density dependent along the star. The main microscopic effect of  $\textbf{B}$ on the boson matter is the splitting of the pressures in two components, one parallel and the other perpendicular to the field direction, giving rise to anisotropic EoS \cite{Angulo:2018url}. In order to take into account this anisotropy, the stable configurations of magnetized BECs were studied using the $\gamma$-structure equations for axially symmetric objects \cite{Samantha,Angulo:2018url}, and supposing pure stars, i.e. that all the matter and the magnetic field in the system are confined inside the compact object and therefore the pressure is zero at its surface. 

We obtained that magnetized BECs are less massive and smaller than the non-magnetic ones, being these effects more relevant at low densities. However, we did not analyze the influence that varying the boundary conditions might have on the solutions of $\gamma$--structure equations. Boundary conditions encloses the physical information about the star surface and surroundings, or about the next layer of matter in the case we deal with a more complex model of a star formed by several shells of different composition. From a mathematical point of view, using one or another boundary condition might lead to different numerical results and even to different qualitative behaviour. For that reason, the goal of this work is to analyze the impact of boundary conditions in the mass--radii curves of magnetized BECs.

The paper is organized as follows. In Section II we summarize the EoS of magnetized BECs for the case of uniform and constant magnetic field, and the $\gamma$--structure equations. In Section III the $\gamma$--structure equations are solved for two boundary conditions and the results are compared. Section IV is devoted to conclusions.  

\section{Magnetized BEC stars}\label{sec2}

A star fully composed of a degenerate gas of interacting neutral vector bosons of mass $m$ and magnetic moment $\kappa$, under the action of a constant and uniform magnetic field, is described by the EoS \cite{Angulo:2018url}
\begin{subequations}\label{EoSRtotal}
	\begin{align}
		P_{\parallel} &= \frac{1}{2}u_0 N^2 -\Omega_{vac} -\frac{B^2}{8 \pi}, \\ P_{\perp} &= \frac{1}{2}u_0 N^2-\Omega_{vac}-B {\mathcal M}+\frac{B^2}{8 \pi},  \\
		{\mathcal M} &= \frac{\kappa}{\sqrt{1-b}} N, \label{EoSRtotal1}\\
		E&=\frac{1}{2}u_0 N^2 +m \sqrt{1-b} N + \Omega_{vac} + \frac{B^2}{8 \pi}.  \label{EoSRtotal3}
	\end{align}
\end{subequations}

Eqs.~\ref{EoSRtotal} contain the dependence of the parallel and perpendicular pressures, $P_{\parallel}$ and $P_{\perp}$, the magnetization $\mathcal M$, and the internal energy $E$ on the particle density $N$ and the magnetic field $b$, with $b=B/B_c$. $B_c = m / 2 \kappa \simeq 10^{19}$~G is the value at which the magnetic energy becomes comparable with the rest energy of the particles.

Due to the assumptions of our model, $m$ and $\kappa$  are twice the neutron mass and magnetic moment respectively. The term $u_0 N^2/2$ that appears on the energy and the pressures stands for the boson-boson interaction  $u_0 = 4 \pi a/m$.  Since $m$ is fixed, the strength of the interaction will be uniquely determine by the scattering length $a$, whose values are in the range of one to few tens of fermi  \cite{Chavanis2012,latifah2014bosons}. In these references it was also shown that a greater $a$ implies stiffer EoS, and consequently heavier and larger stars.

\begin{figure}[h]
	\centering
	\includegraphics[width=0.49 \linewidth]{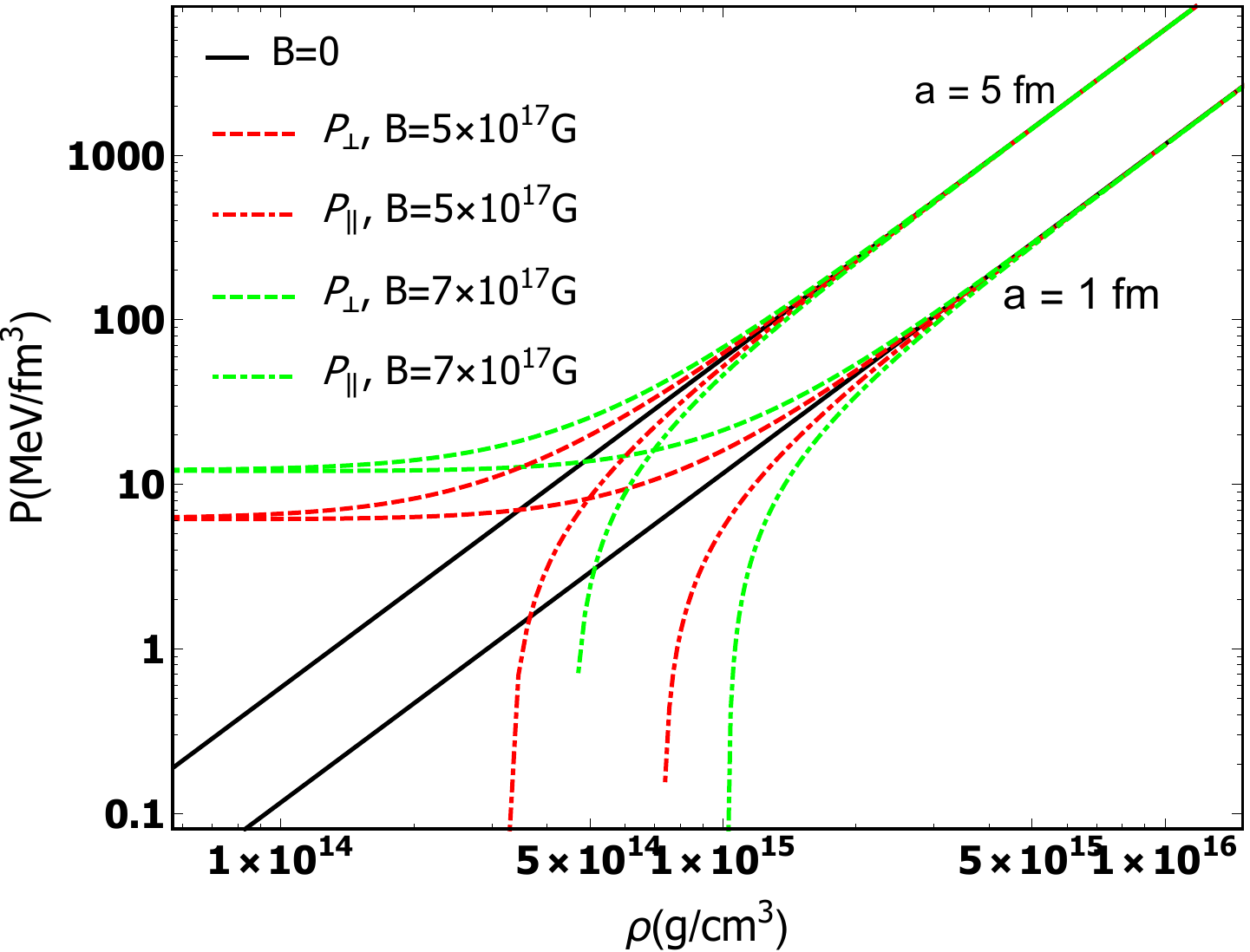}
	\vspace{10pt}
	\includegraphics[width=0.49 \linewidth]{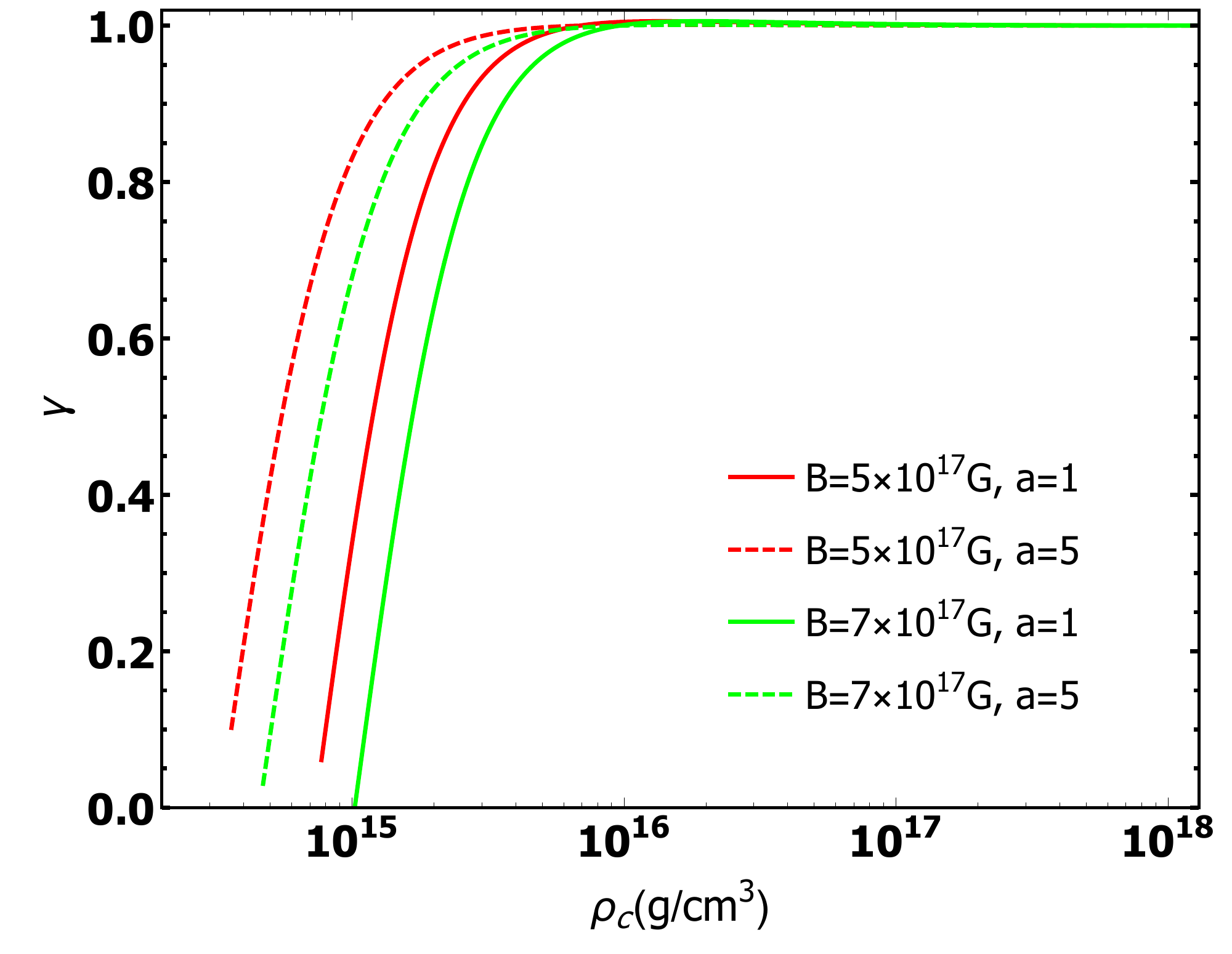}
	\caption{Upper panel: parallel and perpendicular pressures of the BECs as a function of baryon mass density $\rho = m N$, for several values of $a$ and $B$. Lower panel: the parameter $\gamma$ of Eqs.~\ref{gTOV} for the EoS Eqs.~\ref{EoSRtotal} and several values of $a$ and $B$.}
	\label{EoS}
\end{figure}

The terms $B^2/8 \pi$ in Eqs.~\ref{EoSRtotal} accounts for the magnetic field energy and pressure while the quantity $\Omega_{vac}$ depends only on $m$ and $b$ and reads \cite{Quintero2017PRC}
\begin{align}
	\Omega_{vac}=\frac{m^4}{288 \pi} \left \{ b^2(66-5 b^2)-3(6-2b-b^2) (1-b)^2\right.  
	\left.  \log(1-b)-3(6+2b-b^2)(1+b)^2 \log(1+b) \right \}.
\end{align}The anisotropy in the pressure of the magnetized BECs is shown in  the upper panel of Fig.~\ref{EoS} for some typical values of the mass density $\rho = m N$, the parameter $a$ and the magnetic field. At the lower densities, when $\rho$ decreases, $P_{\perp}$ tends to the constant value $B^2/{8 \pi}$, regardless the strength of the interaction (the value of $a$), while $P_{\parallel}$ becomes negative due to the joint contribution of the terms $-\mathcal{M}$ and $-B^2/{8 \pi}$. At a fixed value of the magnetic field, the  increase of the density practically erases the difference between the pressures. However, the anisotropy at low densities  provoke changes in the masses and radii of all the stellar sequence as has been shown in Refs.\cite{Samantha,Angulo:2018url} for BECs and white dwarfs respectively.

To take into account the anisotropy in the pressure when computing the mass-radii curves of the magnetized BECs, in \cite{Samantha} we built an approximated set of structure equations  --the $\gamma$--structure equations-- that describe spheroidal compact objects provided they are not too far from the spherical shape. These equations read
\begin{subequations}\label{gTOV}
	\begin{eqnarray}
	&& \frac{dM}{dr}=r^{2}\frac{(E_{\parallel} +E_{\perp})}{2}\gamma, \label{gTOV1}\\
	&&\frac{dP_{\parallel}}{dz}=-\frac{(E_{\parallel}+P_{\parallel})[\frac{r}{2}+r^{3}P_{\parallel}-\frac{r}{2}(1-\frac{2M}{r})^{\gamma}]}{\gamma r^{2}(1-\frac{2M}{r})^{\gamma}}, \label{gTOV3}\\
	&&\frac{dP_{\perp}}{dr}=-\frac{(E_{\perp}+P_{\perp})[\frac{r}{2}+r^{3}P_{\perp}-\frac{r}{2}(1-\frac{2M}{r})^{\gamma}]}{ r^{2}(1-\frac{2M}{r})^{\gamma}}, \label{gTOV2}
	\end{eqnarray}
\end{subequations}

\noindent where $M(r)$ is the total mass enclosed in the spheroid of equatorial radius $r$ and, at each integration step, $E_{\parallel}$ and $E_{\perp}$ are computed using the parametric dependence of the energy in each pressure derived from Eqs.(\ref{EoSRtotal}). The parameter $\gamma = z/r$ parametrizes the polar radius $z$ in terms of the equatorial one $r$ and  accounts for the axial deformation of the object through the ansatz $\gamma = P_{\parallel c}/P_{\perp c}$ \cite{Samantha}.

Lower panel of Fig.~\ref{EoS} shows $\gamma$
as a function of the star central density for the EoS
depicted in the upper panel. In the region of high density, $\gamma$ is almost $1$ --the star is little deformed-- while as $\rho_c$
decreases, $\gamma$ also diminished, reaching the value $\gamma=0$ when $P_{\parallel}=0$. The decreasing of $\gamma$ is faster when the interaction is weaker. Since $\gamma \simeq 1$ is a requirement to obtain the $\gamma$--structure equations,
in the following we only consider solutions of Eqs.~\ref{gTOV} with $\gamma \geq 0.8$ \cite{Samantha}. According to Eqs.~\ref{EoSRtotal}, for a fixed value of the magnetic field, e.g $B = 5 \times 10^{17}$~G, $\gamma \geq 0.8$  is equivalent to the lower bounds on density $\rho_c \geq 9.32 \times 10^{14}$g/cm$^3$ ($a=5$~fm) and $\rho_c \geq 1.91 \times 10^{15}$g/cm$^3$ ($a=1$~fm). 

To calculate mass-radii curves for a given EoS, Eqs.~\ref{gTOV} are integrated starting from the initial conditions $E_c = E(r=0)$, $P_{\parallel c} = P_\parallel(r=0)$, and $P_{\perp c} = P_\perp(r=0)$, where $E_c$, and $P_{\perp c}$ and $P_{\parallel c}$ are taken from the EoS, until the boundary conditions over the pressures are fulfilled. So far, we have done this process by requiring
\begin{subequations}\label{naked}
	\begin{eqnarray}
	P_{\parallel}(Z)=0,\\
	P_{\perp}(R)=0,
	\end{eqnarray}
\end{subequations}
being $Z$ and $R$ the polar and equatorial radii of the star respectively \cite{Samantha,Angulo:2018url}. These boundary conditions means that we are constructing  a pure compact object, i.e. all the fields that compose the system are contained inside the star. 

However, a usual observed magnitude for magnetic stars is precisely their surface magnetic field. Then, it is to expect that only the matter contribution to the pressures and energy density goes to zero at the surface  of the star, while the magnetic field contribution remains finite. Therefore, to demand that
\begin{subequations}\label{dressed}
	\begin{eqnarray}
	P_{\parallel}(Z)=-\frac{B^2}{8 \pi},\\
	P_{\perp}(R)=\frac{B^2}{8 \pi},
	\end{eqnarray}
\end{subequations}
are more realistic boundary conditions for magnetized compact objects. They mean the star is embedded in what is known as an electro-vacuum, a region without matter where only survive the magnetic field  \cite{Bocquet}. Nevertheless, it is important to remark that this condition is not equivalent to consider a magnetosphere which in general is a more complex structure.

\section{Effects of the boundary conditions on the magnetized BECs}\label{sec3}
\begin{figure}[h]
	\centering
	\includegraphics[width=0.49 \linewidth]{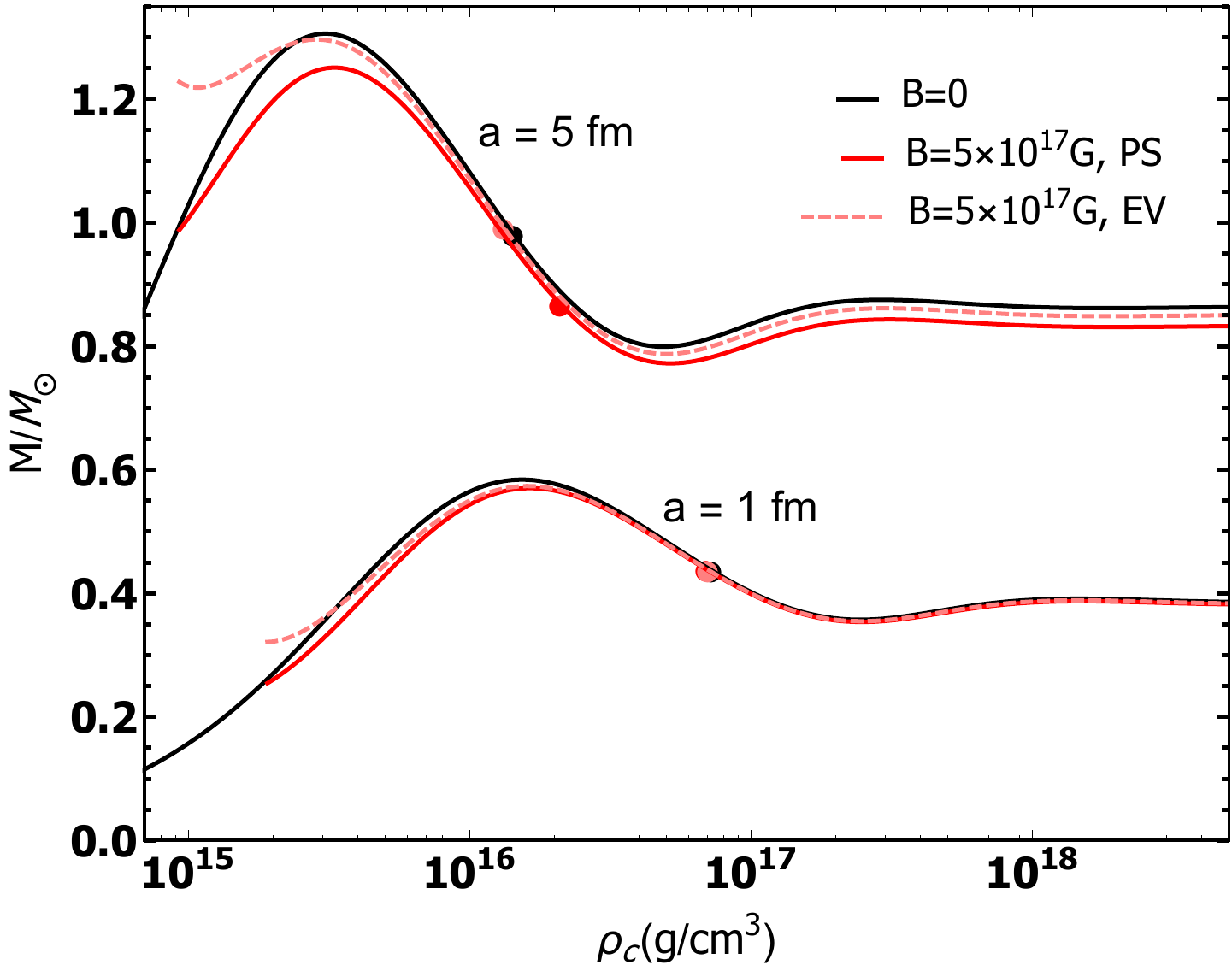}
	\vspace{10pt}
	\includegraphics[width=0.49 \linewidth]{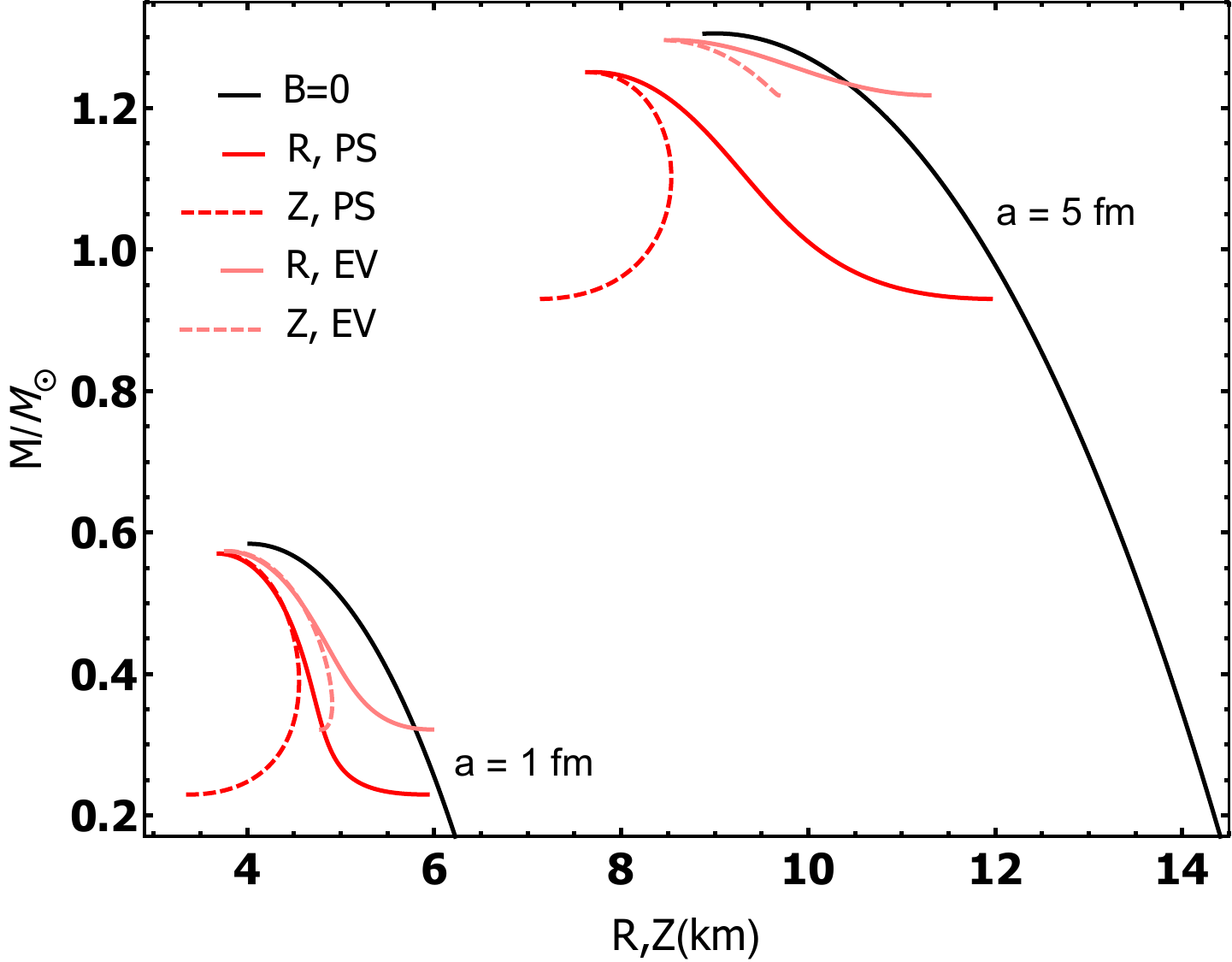}
	\caption{Upper panel: The star masses versus the central mass density for $B=0$ and $B=5 \times 10^{17}$~G with the pure star (PS) and the electrovacuum (EV) boundary conditions. The dots signals the points at which $M = m N$ for each curve.Lower panel: the corresponding mass-radii curves.}
	\label{MR}
\end{figure}

In this section we show the results of integrating Eqs.~\ref{gTOV} with the EoS Eqs.~\ref{EoSRtotal} and the boundary conditions Eqs.~\ref{naked} and \ref{dressed}. We choose $B= 5 \times 10^{17}$~G to do the plots. The $B=0$ curves appears in all the graphics as a reference.  

The curve of the star mass versus the central mass density for the pure star (PS) and the electro-vacuum (EV) boundary conditions have been depicted in upper panel of Fig.~\ref{MR}. The presence of the magnetic field  reduces the star masses \cite{Samantha,Angulo:2018url}, an this reduction is higher in the case of a pure star. But with respect to the $B=0$ curves, the shape of the magnetized stars curves are not so different, beyond the lower bound imposed over the central densities by the condition $\gamma \simeq 1$. Nevertheless, we should keep in mind that not all the stars of the curves depicted in upper panel of Fig.~\ref{MR} are stable solutions of Eqs.~\ref{gTOV}.

On one side, the usual criterion for stability against radial oscillations for TOV--like structure equations requires $dM/dE_{c}>0$ \cite{Samantha}, which for our EoS is equivalent to $dM/d\rho_{c}>0$. Therefore, the stars in the regions where the $M$ vs $\rho$ curves are decreasing should be discarded. On the other hand, for each value of $B$, there is  a central density above which the gravitational mass of the star $M$ becomes larger than its baryon mass $M_B= m \mathcal N$, being $\mathcal N$ the total number of bosons in the star. 
The points at which $M=M_B$ are marked with a solid dot for each curve in upper panel of Fig.~\ref{MR}. At the left of the dots $M< M_B$ and the stars are stable. But at the right $M> M_B$, corresponding to stars configurations that are unstable against its dispersion into free particles at the infinity \cite{PhysRevD.38.2376}. Therefore, this parts of the curves should be also discarded. 

In the lower panel of Fig.~\ref{MR} we show the mass--radii curves for the stable stars of the sequences depicted in the upper panel. As can be appreciated from this plot, the electro-vacuum boundary conditions produce heavier and larger stars than the pure stars ones. These effects increase when the particle--particle interaction grows. 

For $a=1$~fm, when varying from the PS to the EV boundary conditions, the maximum mass increases in a $0.6 \% $ while the corresponding radius augments in a $1.2 \%$ (in this case the anisotropy is negligible and for all the practical purposes the star is spherical). In the case of $a=5$~fm, however, the increase in the maximum mass an its corresponding radius are of a $3 \%$ and a $10 \%$ respectively. 
\begin{figure}[h]
	\centering
	\includegraphics[width=0.5 \linewidth]{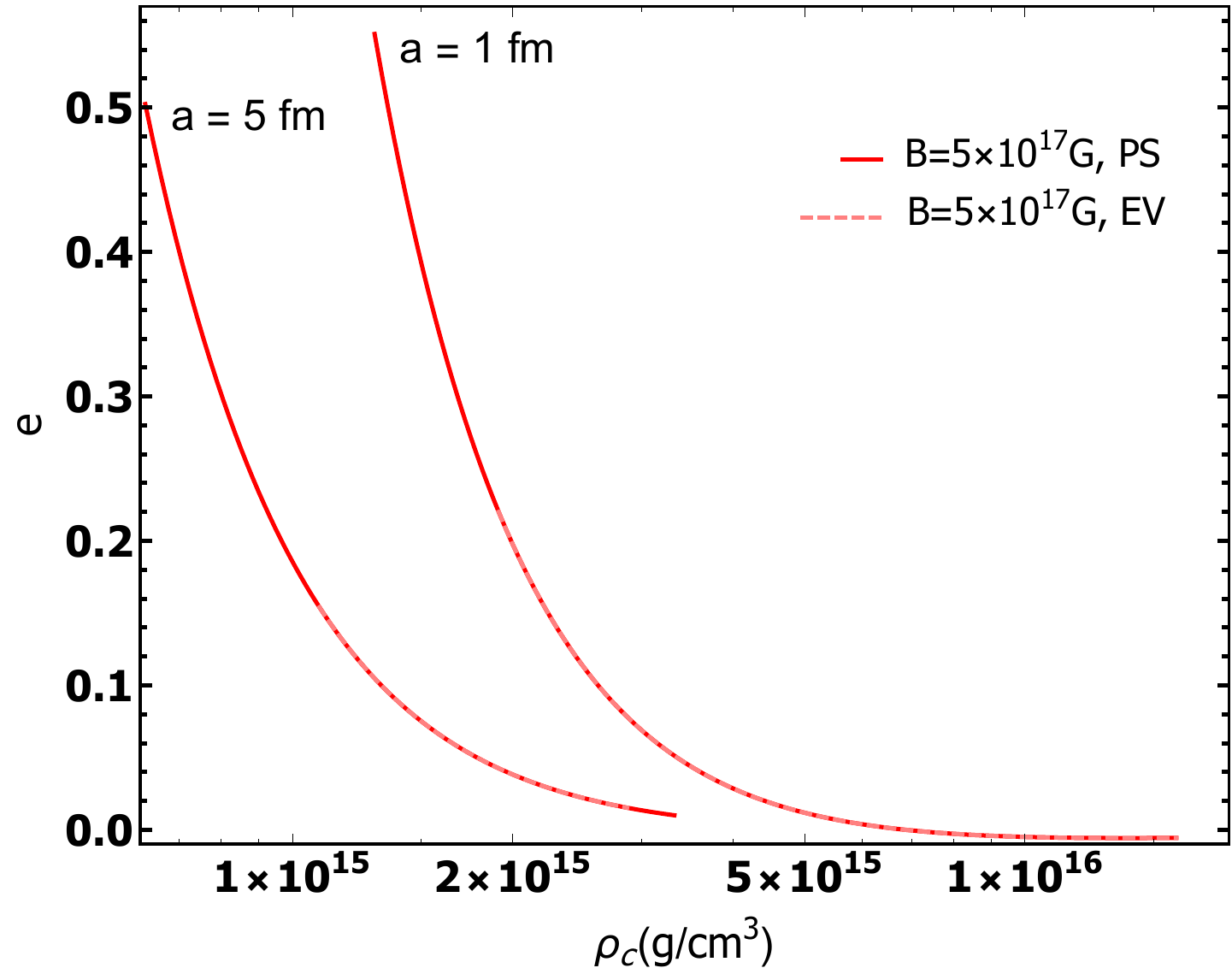}
	\caption{The stable stars ellipticity versus the central mass density for $B=5 \times 10^{17}$~G with the pure star (PS) and the electro-vacuum (EV) boundary conditions.}
	\label{elip}
\end{figure}

On the other hand, as shown in Fig.~\ref{elip}, the boundary conditions does not affect the relatively deformation of the stars. This happens because for the $\gamma$--structure equations $Z = \gamma R$. Therefore the ellipticity $e = (R-Z)/((R+Z)/2)$ \cite{Konno} reduces to $e = 2 (1-\gamma)/(1+\gamma)$, i.e. only depends on the EoS. 

\section{Conclusions}\label{sec5}

We analyzed the impact of the boundary conditions on the solutions of the $\gamma$-structure equations for magnetized Bose--Einstein condensate stars by considering two cases: a pure BECs (PS) and a BECs embedded in a magnetic field (EV). Each of this boundary conditions corresponds to two different physical situations and the use of one or the other modifies the masses and radii of the stars in a way that:

\begin{itemize}
	
	\item The overall behaviour of the mass vs density, and the mass radii curve of the magnetized BECs is not affected.
	
	\item There are less stable stars when using the EV boundary conditions.
	
	\item The use of EV boundary conditions gives heavier and larger stars. Those effect increase with the strength of the interaction, as can be noticed from the plots but also by comparing the maximum masses and the corresponding radii of the stars. 
	
	\item Although the EV stars are bigger than the PS ones, the relative deformation is the same, since for the $\gamma$--structure equations the ellipticity only depend on the EoS.
	
\end{itemize}

Besides of that, as a general conclusion we can state that the observables of the star might be drastically changed when the boundary conditions are varied. Thinking of BECs as an alternative model for NS cores, our result implies that, although in a NS the core accounts for most of their volume and mass, the observables would also depend on the properties of the rest of the layers in a non negligible way.


\section*{Acknowledgments}
The authors thanks Elizabeth Rodr\'iguez Querts for suggesting to carry on this study. G.Q.A, A.P.M and H.P.R have been supported by  PNCB-MES Cuba No. 500.03401 and by the grant of the ICTP Office of External Activities through NT-09.

\bibliography{bibG}

\end{document}